\documentclass[12pt]{article}
 \usepackage{amsmath}
\begin{document}

 \begin{center}
 {\Large \bf Group-theoretic Approach for
 Symbolic Tensor Manipulation: I. Free Indices}
 \end{center}

 \

 \centerline{R. Portugal\footnote{Email: portugal@cbpf.br}}

 \begin{center}
 Laborat\'orio Nacional de Computa\c{c}\~ao Cient\'{\i}fica,\\
 Av. Get\'ulio Vargas, 333,\\
 Petr\'opolis, RJ, Brazil. Cep 25651-070.
 \end{center}

 \

 \centerline{B. F. Svaiter\footnote{Email: benar@impa.br}}

 \begin{center}
 Instituto de Matem\'atica Pura e Aplicada,\\
 Estrada Dona Castorina, 110, \\
 Rio de Janeiro, RJ, Brazil. Cep 22460-320.
 \end{center}

 \

 \begin{abstract}
 We describe how Computational Group Theory provides
 tools for manipulating tensors in explicit
 index notation. In special, we present an
 algorithm that puts tensors
 with free indices
 obeying permutation symmetries into the canonical
 form. The method is based on algorithms
 for determining
 the canonical coset representative of a subgroup
 of the symmetric group. The complexity of our
 algorithm is
 polynomial on the number of indices and
 is useful for implementating general purpose
 tensor packages on the
 computer algebra systems.

 \

 Poisson,
 \end{abstract}

 \

 \

 \section{Introduction}

 The connection between Group Theory and Tensor Calculus was
 established a long time ago. For instance,
 Weyl \cite{Weyl} showed the
 relation of the symmetric group and group rings to
 tensor symmetries and tensor expressions respectively;
 Littlewood \cite{Littlewood} developed tools in
 group representation theory to address the problem of
 determining the dimension of the space generated by tensor
 monomials.
 Recently,
 Fulling et al. \cite{Fulling} used
 these tools to determine the number
 of independent monomials up to order 12 built out of the
 Riemann tensor and its covariant derivative.
 On the other hand, these authors have not established
 a method to determine the independent monomials
 explicitly. The present work performs an important
 step toward this direction.

 The application of Group Theory to develop algorithms
 for tensor manipulation is addressed in refs. \cite{Kryukov}
 and \cite{Dresse}. Ref. \cite{Kryukov} uses group algebra
 which generates an algorithm of exponential
 complexity that can address tensor expressions with at most
 11 or 12 indices.
 Practical applications demand better results.
 Ref. \cite{Dresse}
 uses the backtrack algorithm to find the canonical form of tensor
 expressions built out of totally symmetric or antisymmetric
 tensors.

 The present work addresses the problem of finding efficient algorithms
 for abstract tensor manipulation using Computational Group Theory.
 The main problem consists in simplifying tensor expressions,
 which can be solved if one knows an
 efficient algorithm that puts tensor expressions into
 the canonical form. Ref. \cite{Portugal}
 shows that the problem of finding the canonical
 form of a generic tensor expression reduces to
 finding the canonical forms of single tensors. At this
 point, the group-theoretic approach is the natural
 language to express the problem, since all informations about
 a single tensor can be represented by Group Theory,
 and can be efficiently processed using Computational
 Group Theory.

 We suppose that the tensors have only free indices and
 obey what we now call {\it permutation symmetries}, which are a
 set of tensor equations of the form
 \begin{equation}
 T\,^{{i_{1}}}\,^{\cdots}\,^{{i_{n}}}=\epsilon_\sigma \,T\,^{\sigma (
 {i_{1}}\,\cdots\,{i_{n}})},
 \label{Ts}\end{equation}
 where $\sigma ({i_{1}}\,\cdots\,{i_{n}})$ is a permutation of
 ${i_{1}}\,\cdots\,{i_{n}}$ and
 $\epsilon_{\sigma}$ is either 1 or $-1$. We
 present a polynomial-time algorithm to find the canonical
 forms of these tensors. This algorithm
 is a straightforward extension of algorithms to find
 the canonical coset representative of a subgroup
 of the symmetric group.

 In section 2 we describe the representation theory for tensors.
 In section 3 we present the main algorithm and discuss its
 complexity.


 \section{Representation theory for tensors}

 In this work we use the abstract-index notation for tensor expressions
 such as described in Penrose and Rindler's book
 \cite{Penrose}. We take Lovelock and Rund \cite{Lovelock}
 as a general reference for Tensor Calculus. Ref. \cite{Portugal}
 describes in details how a generic tensor expression can be converted into a sum
 of single tensors. To sum up, a generic
 tensor expression can be expanded so that
 it is a sum of tensor monomials.
 Each monomial is merged into a single tensor that inherits
 the symmetries of the original tensors. If each single tensor can be put
 into the canonical form, then the original tensor expression can be put also.
 At this point the problem consists in finding the
 canonical form of a single tensor obeying permutation symmetries.

 In the present context, we need only three kind of informations
 about a tensor:
 sign, index configuration, and symmetries.
 For example, if one wants to find the canonical form of $T^{cba}$
 knowing that the rank-3 tensor $T$ is totally
 antisymmetric, one starts with
 $$\{+1,[c,b,a]\}$$
 and ends up with
 $$\{-1,[a,b,c]\}.$$
 The natural canonical configuration is $-T^{abc}$.

 In order to use a group-theoretical approach for this kind of
 manipulation we have to represent the symmetry as some group
 which acts on the index configuration and on the sign. This
 goal is achieved in the following way.

 \

 \noindent
 {\bf Definition 1} (Symmetry of rank-$n$ tensor from the
 group-theoretic point of view.)
 Let $S_n$ be the symmetric group on the set
 of points $\{1,2,\cdots,n\}$
 and $H$ the group $(\{+1,-1\},\,\times\,)$
 with multiplicative operation.
 A \textit{tensor symmetry} $S$ is a proper subgroup of the
 external direct product $H\otimes S_n$ such that
 $(-1,{\rm \textit{id}}) \not\in S$,
 where ${\rm \textit{id}}$ is the identity of $S_n$.

 \

 In order to fix notation, from now on we assume that
 $H$ and $S$ are the groups described in Def. 1.
 We also assume that $G$ is the subgroup of $S_n$
 such that $H\otimes G=<(-1,{\rm \textit{id}}),S>$,
 i.e. $H\otimes G$ is the smallest subgroup of $H\otimes S_n$
 which contains $S$ and $(-1,{\rm \textit{id}})$.
 Therefore, the order of $H\otimes G$ is two times
 the order of $S$.

 Now let us describe formally how  $S$ acts on a tensor.
 Each element of $S$ is a pair consisting of
 a sign $(\pm 1)$ and a permutation. The sign of a permutation
 $\pi$ will be denoted by $\epsilon_\pi$. So, an
 element of $S$ has the form $(\epsilon_\pi,\pi)$, where
 $\epsilon_\pi=\pm 1$ and $\pi\in G$.
 The action of $s$ on a totally contravariant rank-$n$ tensor
 with index configuration $T^{i_1 i_2 \cdots i_n}$ is
 \begin{equation}
 s(T^{i_1 i_2 \cdots i_n})=\epsilon_\pi\,
 T^{i_{1^\pi}\, i_{2^\pi} \cdots \, i_{n^\pi}},
 \end{equation}
 where the notation $i_{1^\pi}$ means that the subscript of $i$ is
 the image of point 1 under the action of permutation $\pi$. This
 notation clearly shows that the permutation acts on the
 positions of the indices and seems to be superior to the one
 used in (\ref{Ts}).

 The notation $+\pi$ stands for $(1,\pi)$ and
 $-\pi$ for $(-1,\pi)$.
 For example, the group $S$
 which describes the symmetry of a totally antisymmetric
 rank-3 tensor is
 \begin{equation}
 S=\{{\rm +\textit{id}},\,-(1,2),\,-(1,3),\,-(2,3),\,
 +(1,2,3),\,+(1,3,2)\},
 \end{equation}
 which, in tensor notation, corresponds to
 \begin{eqnarray}
 T^{abc}=T^{abc},\,\,\,\,\,T^{abc}=-T^{bac},\,\,\,
 T^{abc}=-T^{cba}, \nonumber \\
 T^{abc}=-T^{acb},\,\,\,\,\,\,T^{abc}=T^{bca},\,\,\,\,\,\,\,\,
 T^{abc}=T^{cab}.
 \end{eqnarray}
 We use the term ``permutation'' to describe the
 elements of $S$ extending the syntax used
 for elements of $S_n$. The term ``permutation sign''
 refers to $\epsilon_\pi$. The sign $\epsilon_\pi$
 has no relation to the parity of $\pi$ in the general
 case.

 With no loss of generality, we take
 \begin{equation}
 T^{i_1 i_2 \cdots i_n}
 \label{sc}
 \end{equation}
 as the standard configuration. All other configurations
 are obtained by acting permutations of $H\otimes S_n$
 on (\ref{sc}). A generic configuration is
 denoted by $T^{j_1 \cdots j_n}$. We use the sequence
 of $j$'s $({j_1 \cdots j_n})$ as a generic permutation
 of ${i_1 \cdots i_n}$. The standard configuration
 (\ref{sc}) is associated with the element $(+1,id)$, which
 is the minimal element of $H\otimes S_n$.

 In Def. 1, the requirement that $(-1,{\rm \textit{id}})\not\in S$
 avoids the cancelation of a tensor independently
 of its components. Note that if both
 $(+1,\pi)$ and $(-1,\pi)$ are
 in $S$, then we come to two conclusions:
 there are two equal index configurations such that
 $T^{j_1 j_2 \cdots j_n}=-T^{j_1 j_2 \cdots j_n}$
 and $(-1,{\rm \textit{id}})\in S$.
 Vice-versa, if $(-1,{\rm \textit{id}}) \in S$ then
  $(-1,\pi) \in S$ if and only if $(+1,\pi) \in S$.
 When implementing a tensor package in some computer
 language, it is useful to have an efficient
 method to exhibit the symmetries that cancels the
 tensor independently of its components.
 Algorithms for testing membership
 provide such method.

 The index configuration $T^{j_1 j_2 \cdots j_n}$
 is equivalent to  $T^{i_1 i_2 \cdots i_n}$
 if and only if
 there is an element in $S$ such that
 \begin{equation}
 s(T^{i_1 i_2 \cdots i_n}) \equiv T^{j_1 j_2 \cdots j_n}.
 \end{equation}
 In other words, there is an element $(+1,\pi)\in S$ such that
 the lists ${j_1 j_2 \cdots j_n}$ and
 $i_{1^\pi}\, i_{2^\pi} \cdots \, i_{n^\pi}$
 are exactly the same.

 \

 \noindent
 {\bf Proposition 1} The set of index configurations equivalent
 to $T^{i_1 i_2 \cdots i_n}$ is given by the action of $S$ over
 $T^{i_1 i_2 \cdots i_n}$. The cardinality of this set is
 the order of $G$ ($|G|$).

 \

 \noindent
 Note that $|G|=|S|$.

 \

 \noindent
 {\bf Proposition 2} Let $T^{i_{1^\pi}\, i_{2^\pi} \cdots \, i_{n^\pi}}$ be
 an index configuration which is not equivalent to
 $\pm T^{i_1 i_2 \cdots i_n}$ (i.e. both $+\pi)$ and $-\pi)$
 are not in $S$).
 The set of configurations
 equivalent to $T^{i_{1^\pi}\, i_{2^\pi} \cdots \, i_{n^\pi}}$ is
 the right coset of $S$ in $H\otimes S_n$ which contains $(+1,\pi)$.
 The number of index configurations equivalent to
 $T^{i_{1^\pi}\, i_{2^\pi} \cdots \, i_{n^\pi}}$ is $|G|$.

 \

 \noindent The final part of proposition 2 follows from
 the fact that all cosets of $S$ in
 $H\otimes S_n$ have cardinality $|G|$.

 \

 \noindent
 {\bf Proposition 3}
 Consider a set of independent index configurations with positive
 coefficients. The maximum cardinality of this
 set is the index
 of $G$ in $S_n$ ($|S_n$:$G|$).

 \

 Usually, no one describes the symmetry of a tensor in
 index notation by listing all equivalent index
 configurations. In general, one
 gives a few equations and skips the ones that can be obtained
 from the original equations.
 The following example illustrates this situation.
 Let $T^{abcd}$ be a rank-4 tensor with the
 following permutation symmetries.
 \begin{subequations}
 \begin{align}
 T^{abcd}=-T^{bacd},
 \label{s1} & \\
 T^{abcd}=T^{cdab}.
 \label{s2} &
 \end{align}\label{s1s2}
 \end{subequations}

 \noindent
 >From these equations we can obtain
 \begin{equation}
 T^{abcd}=-T^{abdc}.
 \label{s3}
 \end{equation}
 Note that one can describe the symmetries of $T^{abcd}$
 by using a different
 set of equations, for example eqs. (\ref{s2}) and (\ref{s3}).
 In group-theoretic language, this is
 equivalent to describe a group by generators. In the example
 above, the generating set $K$ for the symmetry of $T^{abcd}$,
 described by eqs. (\ref{s1s2}), is
 \begin{equation}
 K=\{-(1,2),\,+(1,3)(2,4)\}.
 \label{K}
 \end{equation}
 The symmetry $S$ is the group generated by $K$ ($<$$K$$>$)
 \begin{eqnarray}
 S=\{+{\rm \textit{id}},\,-(1,2),\,-(3,4),\,+(1,2)(3,4),
 \,+(1,3)(2,4),\, \nonumber \\
 -(1,3\,2,4),\,-(1,4,2,3),\,+(1,4)(2,3)\}.
 \label{S1}
 \end{eqnarray}
 The group $G$ is generated by $\{(1,2),\,(1,3)(2,4)\}$ and is
 obtained from (\ref{S1}) by removing the permutation signs.
 >From proposition 1, the set of configurations equivalent to
 the standard configuration
 $T^{abcd}$ is given by the action of all elements of $S$ on
 $T^{abcd}$, which yields
 \begin{equation}
 \{T^{abcd},-T^{bacd},-T^{abdc},T^{badc},
 T^{cdab},-T^{cdba},-T^{dcab},T^{dcba}\}.
 \label{Tabcd}
 \end{equation}
 The number of equivalent configurations is 8, the order of $S$.
 Now consider $T^{acbd}$, which is not in set (\ref{Tabcd}).
 This index configuration is obtained from $T^{abcd}$ by the
 action of $+(2,3)$. Neither $+(2,3)$ nor $-(2,3)$
 are in $S$.
 Proposition 2 says that the set of configurations equivalent
 to $T^{acbd}$ is the right coset of $S$ in $H\otimes S_4$
 which contains $+(2,3)$. This coset is obtained by multiplying
 each element of $S$ by $+(2,3)$:
 \begin{eqnarray}
 S\times (+(2,3)) =\{+(2,3),\,-(1,3,2),\,-(2,3,4),\,
 +(1,3,4,2), \nonumber \\
 \,+(1,2,4,3),\,-(1,2,4),\,-(1,4,3),\,+(1,4)\}.
 \end{eqnarray}
 The action of this coset on $T^{abcd}$ generates all index
 configurations equivalent to $T^{acbd}$.
 Note that $|S_4$:$G|=3$. From proposition 3 we know that the set
 of independent configurations has cardinality 3.
 An example of this set is
 \begin{equation}
 \{T^{abcd},T^{acdb},T^{adcb}\},
 \label{Tabcd3}
 \end{equation}
 which is a complete right transversal for $G$ in $S_4$
 in tensor notation.
 One can easily recognize that eqs. (\ref{s1s2}) and
 (\ref{s3}) represent the symmetries of the Riemann
 tensor without taking into account the cyclic symmetry.
 The group generated by $\{(1,2),\,(1,3)(2,4)\}$ is the
 dihedral group of order 8 ($D_8$).
 Then, the symmetry of the Riemann
 tensor given by (\ref{S1})
 is the largest subgroup of $H\otimes D_8$
 which does not contain $(-1,id)$.



 In order to address the problem of finding the canonical
 configuration equivalent to a given index configuration,
 we have to define an order for the permutations of $S_n$.
 Let $\boldsymbol b=[b_1,\cdots,b_n]$ be a list of $n$ distinct
 points of the set $\{1,\cdots,n\}$.
 Suppose that $p_k$ and $p_l$ are points. We
 define the order ``$\prec_{\boldsymbol b}$''
 for points with respect to $\boldsymbol b$ in the following way:
 $p_k \prec p_l$ if $k<l$. So,
 $p_k$ is smaller that $p_l$ if $p_k$ comes before $p_l$ in
 $\boldsymbol b$.
 We omit the index $\boldsymbol b$ from
 ``$\prec_{\boldsymbol b}$'' to simplify the notation.
 $\boldsymbol b^\pi=[b_1^\pi,\cdots,b_n^\pi]$
 is the image of $\boldsymbol b$ under $\pi$.
 Define
 \begin{equation}
 \mathcal{L}=\{\boldsymbol b^\pi,\,\pi\in S_n\}.
 \end{equation}
 Now let extend the order ``$\prec$'' to the elements of $\mathcal{L}$.
 Let $L_1$ and $L_2$ be in $\mathcal{L}$.
 $L_1 \prec L_2$ if
 \begin{equation}
 L_1[1] \prec L_2[1] \,\,\,\,\, {\rm or} \,\,\,\,\, (L_1[1]=L_2[1]
 \,\,\,\, {\rm and}
  \,\,\,\, L_1[2..n] \prec L_2[2..n]),
 \end{equation}
 where $L[i]$ means the $i$-th element of $L$ and
 $L[2..n]$ means $\{L[2],L[3],\cdots,L[n]\}$.
 If $\pi_1$ and $\pi_2$ are in $S_n$ then
 \begin{equation}
 \pi_1 \prec \pi_2 \Leftrightarrow \boldsymbol b^{\pi_1}
 \prec \boldsymbol b^{\pi_2}.
 \end{equation}
 Now we extend the order ``$\prec$'' to group $S$:
 \begin{equation}
 (\epsilon_{\pi_1},\pi_1) \prec
 (\epsilon_{\pi_2},\pi_2) \Leftrightarrow \pi_1 \prec \pi_2.
 \label{order2}
 \end{equation}
 Recall that $(+1,\pi)$ and $(-1,\pi)$ cannot be at the
 same time in $S$. So we simply disregard the sign as stated in
 (\ref{order2}).  ``$\prec$'' is a well-order
 (total order with a minimal element)
 in $S_n$, $S$, and in any
 coset of $S$ in $H\otimes S_n$.
 >From now on, ``minimal point'' refers to
 the order ``$\prec$'' with respect to some
 $\boldsymbol b$.

 A \textit{canonical right transversal}
 for $G$ in $S_n$ ($S$ in $H\otimes S_n$)
 is a complete right transversal for $G$ in $S_n$
 ($S$ in $H\otimes S_n$) such that
 each coset representative is minimal.
 The set (\ref{Tabcd3}) is the canonical right transversal
 in tensor notation for group $G$ in $S_4$ with respect
 to $\boldsymbol b=[1,3,2,4]$, where $G$ is
 generated by $\{(1,2),\,(1,3)(2,4)\}$.
 Note that the order ``$\prec$'' allows to sort
 a canonical right transversal for $G$ in $S_n$. This
 is important for addressing the simplification problem
 when there are side relations, such as the cyclic symmetry.
 This kind of symmetry is not addressed in this paper.

 \section{Algorithm to canonicalize tensors with free indices}

 Now we address the following problem.
 Suppose one gives a totally contravariant tensor with
 the symmetries described by a set of tensor equations,
 and a free index configuration that is not the standard
 one.
 Find the canonical index configuration with respect to
 the order ``$\prec$''.
 For example, suppose that rank-4 tensor $T$ has the
 symmetries
 (\ref{s1s2}), and one gives the following index
 configuration: $T^{bcad}$. What is the canonical
 configuration with respect to
 $\boldsymbol b=[1,3,2,4]$?

 Using the representation theory, the problem above
 can be solved if one knows the solution of the following
 problem. Given a generating set $K$ for the group $S$
 and an element $(\epsilon_\pi,\pi)$ in $H\otimes S_n$,
 find the canonical coset representative of the
 coset $S\times (\epsilon_\pi,\pi)$
 with respect to the order defined by (\ref{order2}).

 The answer to this
 problem is an algorithm that can be used to put
 a tensor with free indices into the canonical form.
 Before describing the algorithm, we extend
 some well known definitions for permutations groups,
 in order to use them in connection with $S$, which is
 a direct product of groups. The extensions
 are straightforward.

 \

 \noindent
 {\bf Definition 2} Let $p$ be a point in the set
 $\{1,\cdots,n\}$.
 The \textit{stabilizer of p in S} is the subgroup
 $S_{p}$ defined by
 \begin{equation}
 S_{p}=\{s\in S \, | \, p^s=p\},
 \end{equation}
 where $p^s=p^\pi$ and $s=(\epsilon_{\pi},\pi)$.

 \

 \noindent
 In other words, $S_{p}$ consists of all elements of $S$
 that fix the point $p$. $S_{p}$ is a subgroup of $S$.
 If $s=(\epsilon_{\pi},\pi)\in S$, ``$s$ fixes $p$'' means
 ``$\pi$ fixes $p$''.

 \

 \noindent
 {\bf Definition 3} Let $Q$ be a subset of the set
 of points $\{1,\cdots,n\}$.
 The \textit{pointwise stabilizer of $Q$ in S} is the subgroup
 $S_{Q}$ defined by
 \begin{equation}
 S_{Q}=\{s\in S \, | \, \forall q\in Q,\,q^s=q\}.
 \end{equation}
 In other words, $S_{Q}$ fixes all points of the subset $Q$.

 \

 \noindent
 {\bf Definition 4} A ordered subset
 $\boldsymbol b=[b_1,\cdots,b_m]$ of the set of
 points $\{1,\cdots,n\}$ is a \textit{base} for $S$
 if $S_{b_1,\cdots,b_m}=(1,{\rm \textit{id}})$.

 \

 \noindent
 This means that the only element of $S$ that fixes all points
 of $\boldsymbol b$ is the identity. A useful property of a base is that an
 element of $S$ is uniquely determined by the base image.
 We can order the points $\{1,\cdots,n\}$ such that
 the base points are the first $m$ points. Let us name the remaining  points
 $l_1,\cdots,l_{n-m}$. They simply follow the usual
 increasing order. From now on, the order ``$\prec$'' defined
 in the previous section is based on the set
 $[b_1,\cdots,b_m,l_1,\cdots,l_{n-m}]$.

 \

 \noindent
 {\bf Definition 5} A \textit{strong generating set K for S}
 relative to the base $\boldsymbol b=[b_1,\cdots,b_m]$
 is a generating set for $S$ with the following property:
 $K \cap S_{b_1,\cdots,b_j}$ is a generating set
 for the pointwise stabilizer $S_{b_1,\cdots,b_j}$, for
 $1\le j \le m-1$.

 \

 \noindent
 In other words, a generating set $K$ for $S$ is \textit{strong}
 if, after selecting the permutations of $K$ that
 fix the points $b_1,\cdots,b_j$, one has a set that
 generates the group $S_{b_1,\cdots,b_j}$. This must be
 valid for $j$ from 1 to $m-1$.

 \

 The input of the algorithm is\\
 (a) an index configuration $T^{j_1 j_2 \cdots j_n}$; and\\
 (b) a base and a strong generating set for $S$.\\
 The output is the canonical index configuration, which
 is obtained by the action of the canonical coset
 representative on the standard
 configuration $T^{i_1 i_2 \cdots i_n}$.
 If $S$ is described by a generating set that is not strong,
 the first step is to obtain a strong generating set
 and a corresponding base. Refs. \cite{Leon} and \cite{Butler_book}
 present algorithms that perform this task. These algorithms
 must be extended in order to work within $H\otimes S_n$.

 The algorithm Canonical described below uses the general structure
 of \textit{chain of stabilizers} developed by Sims \cite{Sims}
 and is an straightforward modification of the algorithm presented
 by Butler \cite{Butler}. Following Sim's notation, let
 $S^{(i)}$ be the group that stabilizes the points
 $\{b_1,\cdots,b_{i-1}\}$,
 i.e. $S^{(i)}=S_{b_1,\cdots,b_{i-1}}$. Then $S^{(1)}=S$ and
 $S^{(m)}=(1,id)$. Let $\Delta^{(i)}$ be the orbit
 of $S^{(i)}$ that contains the point $b_i$, i.e.
 $\Delta^{(i)}=b_i^{S^{(i)}}$. Suppose that $K$ is a
 strong generating set for $S$ and define
 $K^{(i)}=K\cap S^{(i)}$. $K^{(i)}$ is a strong generating
 set for $S^{(i)}$, for $1 \le i \le m$.
 Let  $\nu^{(i)}$ be the Schreier vector of
 $\Delta^{(i)}$ with respect to generators $K^{(i)}$. Here, the
 components of the Schreier vectors are elements of $S$. Following
 Butler \cite{Butler_book}, if $q\in \Delta^{(i)}$, let
 $trace(q,\nu^{(i)})$ be the element
 $(\epsilon_\omega,\omega)$ of $S^{(i)}$
 such that $p^\omega=q$, where $p$ is the minimal point of
 the subsets of
 $\Delta^{(i)}$ that contains $q$.

 The algorithm consists of $m$ loops. Suppose that
 $[q_1,\cdots,q_m]$ is the base image of the canonical
 representative. The $i$-th loop finds permutation
 $(\epsilon_\lambda,\lambda)$ that determines $q_i$,
 i.e. $b_i^\lambda=q_i$. The permutation
 $(\epsilon_\lambda,\lambda)$ obeys the
 constraint
 \begin{equation}
 [(b_1)\,^\lambda,\cdots,(b_{i-1})^\lambda]=
 [q_1,\cdots,q_{i-1}].
 \label{cons}
 \end{equation}
 The set of all elements of $S$ that obey
 (\ref{cons}) is given by
 $S^{(i)}\times (\epsilon_\lambda,\lambda)$,
 then $b_i^{S^{(i)}\times (\epsilon_\lambda,\lambda)}$
 yields all possible images of $b_i$ in the coset
 $S^{(i)}\times (\epsilon_\lambda,\lambda)$.
 $q_i$ is the minimal point of these images.
 At the last loop, $(\epsilon_\lambda,\lambda)$
 gives the complete base image
 $[q_1,\cdots,q_m]$.

 \

 \

 \

 \centerline{\bf Algorithm Canonical (free indices)}

 \begin{tabbing}
 {\bf Input:} \= $T^{j_1 \cdots j_n}=
 \epsilon_\pi T^{i_{1^\pi}\, i_{2^\pi} \cdots \, i_{n^\pi}}$, where
 $(\epsilon_\pi,\pi)\in S$; \\
 \> $\boldsymbol b$ \= $=[b_1,\cdots,b_m]$ base for $S$; and \\
 \> $K_S$ strong generating set for $S$ with respect to $\boldsymbol b$.\\
 $K^{(i)}$.\\
 \\
 {\bf Output:} \= $\epsilon_\lambda T^{i_{1^\lambda}\, i_{2^\lambda} \cdots
 \, i_{n^\lambda}}$,
 where
 $(\epsilon_\lambda,\lambda)$ is the canonical representative of the\\
 \>  coset of $S$ in  $H\otimes S_n$ that contains the permutation
 $(\epsilon_\pi,\pi)$.\\
 \\
 {\bf be}\={\bf gin}\= \\
 \> $(*$ initialization of $(\epsilon_\lambda,\lambda)$ and $K$ $*)$\\
 \> $(\epsilon_\lambda,\lambda)\,:=\,(\epsilon_\pi,\pi)$; \\
 \> $K\,:=\,K_S$;\\
 \\
 \> {\bf for} $i$ from 1 to $m$ {\bf do} \\
 \> \> $(*\,\Delta$ is the basic orbit $\Delta^{(i)}\, *)$\\
 \> \> $\Delta\,:=\,b_i^{<K>}$;\\
 \> \> $k\,:=\,$ position of the minimal point of $\Delta^\lambda$;\\
 \> \> $p\,:=\,k$-th point of $\Delta$;\\
 \> \> $(\epsilon_{\omega},\omega)\,:=\,trace(p,\nu)$,
 where $\nu$ is the Schreier vector of $\Delta$ with respect to $K$; \\
 \> \> $(\epsilon_{\lambda},\lambda)\,:=\,(\epsilon_{\omega},
 \omega)\times (\epsilon_{\lambda},\lambda)$; \\
 \> \> $K\,:=\,$remove permutations of $K$ that have point $b_i$; \\
 \> {\bf end for}; \\
 \\
 \> {\bf return} $\epsilon_\lambda T^{i_{1^\lambda}\, i_{2^\lambda} \cdots
 \, i_{n^\lambda}}$;
 \\
 {\bf end} \\
 \end{tabbing}

 Let us see an example. If one gives the index
 configuration $T^{bcad}$, where $T$ has the symmetries
 (\ref{s1s2}), the element $(\epsilon^\pi,\pi)$ is
 $+(1,2,3)$. A base for $S$ is $[1,3]$, so the
 order ``$\prec$'' is based on the list
 $\boldsymbol b=[1,3,2,4]$. A strong generating set
 with respect to base  $[1,3]$ is
 \begin{equation}
 K=\{-(1,2),\,-(3,4),\,+(1,3)(2,4)\}.
 \label{strongK}
 \end{equation}
 First loop yields: $\Delta:=\{1,2,3,4\}$;
 $\Delta^{(1,2,3)}:=\{2,3,1,4\};$
 $k:=3;$ $p:=3;$
 $(\epsilon_\omega,\omega):=+(1,3)(2,4)$, since
 $b_1^\omega=p;$
 $(\epsilon_\lambda,\lambda):=+(2,4,3);$
 $K:=\{-(3,4)\}$.
 $(\epsilon_\lambda,\lambda)$ applied on
 $T^{abcd}$ gives $T^{adbc}$. Second
 loop yields: $\Delta:=\{3,4\}$;
 $\Delta^{(2,4,3)}:=\{2,3\};$
 $k:=2$, since $3\prec 2$; $p:=4;$
 $(\epsilon_\omega,\omega):=-(3,4);$
 $(\epsilon_\lambda,\lambda):=-(2,4);$
 $K:=\{\,\}$. The algorithm finishes and
 the canonical configuration is
 $\epsilon_\lambda\,
 T^{i_{1^\lambda}i_{2^\lambda}i_{3^\lambda}i_{4^\lambda}}=
 -T^{adcb}$.

 The algorithms to find strong generating set,
 basic orbit, Schreier vector, and \textit{trace}
 for permutation groups
 are described in refs. \cite{Leon} and
 \cite{Butler_book}. The extension
 of these algorithms to work within the direct product $H\otimes S_n$
 is straightforward if one uses the fact that a image
 of a point $p$ under the action of $(\epsilon_\pi,\pi)$ is $p^\pi$.
 The product of permutations in $S_n$ is naturally
 extended to the product of
 elements of $H\otimes S_n$.

 The analysis of the complexity of the algorithm Canonical is the
 following. Algorithms to find basic orbit, Schreier vector and
 \textit{trace} have an $O(n^2)$ bound.
 Since the algorithm Canonical performs $n$ loops
 in the worst case, it is bounded by $O(n^3)$.
 A strong generating set for symmetry $S$ is required.
 It is known that Schreier-Sims algorithm has
 an $O(n^5)$ bound. So, if the generating set of $S$ is not strong,
 the overall bound for the algorithm to find
 the canonical form of tensors is $O(n^5)$, where $n$ is the number
 of indices.

 %
 %
 %

 \

 \noindent {\bf Acknowledgments}

 \

 \noindent
 We thank Drs. S. Watt and J. Ja\'en for
 stimulating discussions on this subject and
 Dr. M. Rybowicz for providing useful references.

 \end{document}